\documentclass[twocolumn]{revtex4}
\usepackage[colorlinks,bookmarksopen,bookmarksnumbered,citecolor=magenta,urlcolor=red]{hyperref}
\usepackage{graphicx}
\usepackage{amsmath}
\usepackage{natbib}
\begin{document}


\title{Quasi-geostrophic dynamics in the presence of moisture gradients}

\author{Joy M. Monteiro and Jai Sukhatme}

\affiliation{Centre for Atmospheric and Oceanic Sciences, Indian Institute of Science,\\
Bangalore 560012, India} 


\begin{abstract}

The derivation of a quasi-geostrophic (QG) system from the rotating shallow water
equations on a midlatitude $\beta$-plane coupled with moisture is
presented. Condensation is prescribed to occur whenever the moisture at a point
exceeds a prescribed saturation value. It is seen that a slow condensation time scale is required to
obtain a consistent set of equations at leading order. 
Further, since the advecting wind fields are geostrophic, changes in
moisture (and hence, precipitation) occur only via non-divergent mechanisms.
Following observations, a saturation profile with gradients in the zonal and
meridional directions is prescribed. A purely meridional gradient has the effect
of slowing down the dry Rossby waves, through a reduction in the ``equivalent gradient" 
of the background potential vorticity.
A large scale unstable moist mode
results on the inclusion of a zonal gradient by itself, or in conjunction with a meridional moisture gradient. For gradients that are
are representative of the atmosphere, the most
unstable moist mode propagates zonally in the direction of increasing moisture, 
matures over an intraseasonal timescale and has small phase speed.
\vskip 0.2 truecm
\begin{center}
{\bf Journal Ref: QJRMS, DOI:10.1002/qj.2644, 2015.}
\end{center}

\end{abstract}
\vskip 0.25truecm


\maketitle

\section{Introduction}

Following \cite{gill1982}, the study of idealised dynamical models 
forced by active,
condensable scalars has received considerable attention, mainly due to the fact
that these models represent the action of water vapour on the atmosphere in a
simple manner. 
From a tropical perspective, the use of one or two mode shallow water systems on an
equatorial $\beta$-plane is prevalent, 
and a large body of work had been directed towards understanding the so-called convectively 
coupled equatorial waves, i.e., to identify the effects of moisture on ``dry" equatorial waves \citep[see for example, the
recent review by][]{kiladisCCEW}. The vertical decomposition into modes has been done
    in many ways: for example, using analytic functions that satisfy appropriate boundary conditions
    \citep[]{majdaMultiscale2012}, basis functions derived from observations
    \citep[]{neelinQuasiequilibrium2000}. 
Another avenue has been to try and see if the addition of moisture gives rise to modes that do not
exist in the traditional ``dry" equations \citep[see, for example,][]{sobel2001,raymond2007}. 

Outside the tropics, the influence of condensational heating via a moisture
equation on the linear evolution of a quasi-geostrophic (QG) system
\citep{bannon1986}, barotropic \citep{lambaerts2011} and baroclinic instability
\citep{fantiniMoistBaroclinic1993,lambaerts2012}, the life-cycle of baroclinic eddies 
\citep{gutowskiLifeCycles1992} and homogeneous QG turbulence
\citep{lapeyreHeld2004} has also been considered. More pertinent to our work has been
the study of ``diabatic Rossby waves'' which are generated purely due to PV anomalies rsulting from
latent heat release \citep[]{craigCumulus1988, snyderQuasigeostrophic1991, parkerConditional1995}.
Interestingly, applications to other planetary
atmospheres, such as Mars with CO$_2$ condensation and Titan with CH$_4$ condensation, have also been
successfully pursued \citep{sabato2008, mitchell2011}.

\begin{figure}
\centering
\includegraphics[width=0.45\textwidth]{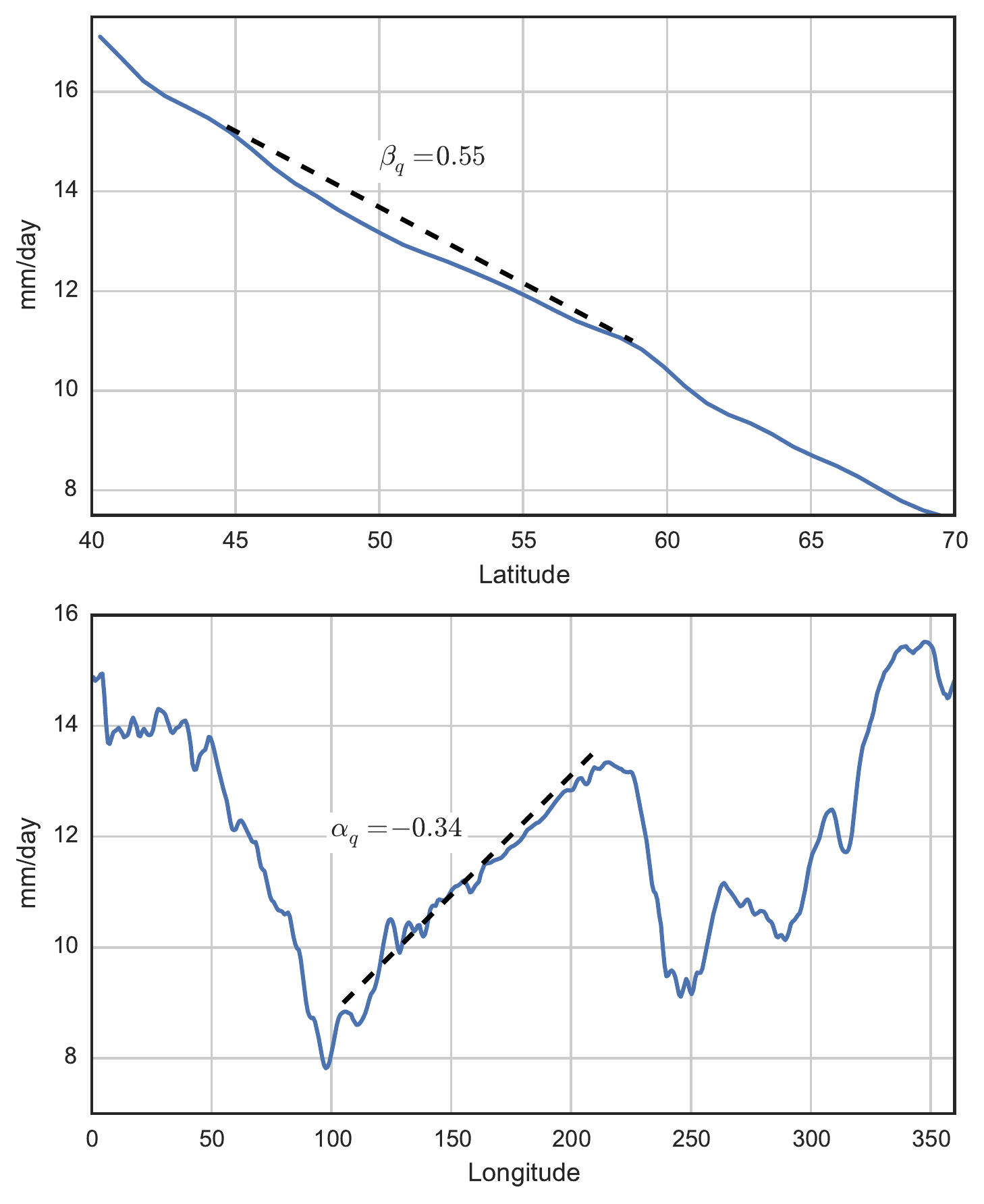}
\caption{Panel \textbf{(a)} shows the meridional profile of the zonally
    averaged climatological precipitable water in the ERA-Interim dataset between 40-70N. This corresponds to a
    non-dimensional $\beta_q$ (see Section 2.3) of about 0.55. Panel \textbf{(b)}
shows the zonal profile of precipitable water averaged between 40-70N. A
strong gradient is seen in both the Pacific and Atlantic oceans,
corresponding to a non-dimensional $\alpha_q$ of about -0.34 (see Section 2.3).} 
\label{pwatGradients}
\end{figure}

In the context of QG dynamics, forcing due to condensation
is usually directly appended to the ``dry" system which makes
it difficult to understand the constraints (if any) that the forcing must obey to be consistent with 
the requirements of QG theory.
Furthermore, most previous studies prescribe a homogeneous background
saturation profile which does not include horizontal gradients. A noteworthy
exception is \cite{sobel2001}; by allowing a meridional gradient in the saturation
profile, their linear analysis on an $f$-plane
showed the emergence of a remarkable unstable eastward propagating mode 
that owed its existence to the aforementioned meridional gradient. As
meridional and zonal gradients in precipitable water are robust
features of the 
atmosphere both in the mid-latitudes (Figure \ref{pwatGradients}) and
the tropics \citep{sukhatme2012}, it is natural to incorporate them along with
a gradient in the planetary vorticity when formulating simple models to
study the effects of moisture on large scale dynamics. In fact, following \cite{sobel2001}, a numerical calculation with
zonal and meridional moisture gradients was
carried out on the equatorial $\beta$-plane by \cite{sukhatmeMoistModes2013}.
However, the complexity of the full shallow water system did not permit a complete 
understanding of the mechanism behind the observed
instability.

In this paper, a QG system is derived from the moist shallow water
equations (SWE) on a midlatitude $\beta$-plane. This clearly delineates the constraints that the
moist forcing must satisfy in order to maintain geostrophy at
leading order. We then proceed to a linear
analysis in the presence of both meridional and zonal gradients of the
saturation profile. With only a meridional gradient, there are no unstable modes and the Rossby waves that emerge are slower than their dry
counterparts. Interestingly,
the slowing down of the waves is traced to a reduction in the ``equivalent gradient" of the background 
potential vorticity. It is seen that in the presence of zonal gradients, the system
supports a large scale unstable moist mode. In fact, with both meridional and
zonal gradients in the saturation profile, with values similar to those observed
in the atmosphere, the most unstable mode propagates zonally in a direction
of increasing moisture, matures over an intraseasonal timescale 
and is characterised by a small phase speed.
The eigenvector structure corresponding to
the waves in the presence of gradients is also examined to determine
the phase relationship between moistening and precipitation.
Finally, a summary of results and a brief discussion conclude the paper.

\section{Derivation of the model}

In deriving our model, we follow \cite{bouchut2009} and consider the one-layer forced SWE. 
In this situation, the forcing due to 
condensation manifests itself as a 
mass source/sink \citep{gill1979}, and the equations read 
\begin{align}
\frac{Du}{Dt} - fv = -g h_x, \nonumber \\
\frac{Dv}{Dt} + fu = -g h_y, \nonumber \\
\frac{Dh}{Dt} + h(\nabla .\mathbf{u}) = -\chi S, \nonumber \\
\frac{Dq}{Dt} + q(\nabla .\mathbf{u}) = -S.
\label{a1}
\end{align}
Here $u,v$ are the horizontal velocities, $h$ is the
height of the shallow water system, $f = f_0 + \beta y$
is the Coriolis parameter and $q$ is the column water vapour.
$S = \delta(q - q_s)/\tau_q$ is a Betts-Miller like condensation
sink for the water vapour equation, where $\tau_q$ is the condensation timescale and $\delta$ is one when
$q > q_s$ and zero otherwise. 
$q_s$ is the (prescribed)
background saturation water vapour
profile. 
With a ``conversion factor" $\chi$ this
takes the form of a mass source/sink. 
We note that this particular parameterisation of condensation is reminiscent of the model
developed by \cite{muller2009}, where precipitation is a
function of column water vapour rather than the saturation vapour
pressure \citep[see also the discussion in][]{holloway-neelin}. 
Writing $q = q_s + \tilde{q}$, the source term
reduces to $S = \delta \tilde{q}/\tau_q$, and the vapour conservation law takes the form:
\begin{equation}
\frac{D\tilde{q}}{Dt} + \tilde{q}(\nabla .\mathbf{u}) + \nabla \cdot (\mathbf{u} q_s) =
-\frac{\delta}{\tau_q} \tilde{q}.
\label{new1}
\end{equation}

where $q_s = q_s(x,y)$ is a function of space only, whereas $\tilde{q} = \tilde{q}(x,y,t)$ is also a
function of time.
\subsection{Conservation Laws}
The ``dry'' shallow water equations respect the material invariance of 
potential vorticity $(f + \zeta)/h$, as well as the global conservation of 
energy $ E = (gh^2 + h\mathbf{u.u})/2$. Eliminating $S$ from the last two
equations in Eq. \ref{a1} to give a single conservation law, the equivalent statements
for (\ref{a1}) are:

\begin{align}
\frac{D}{Dt}(\frac{f+\zeta}{m}) = 0, \\
\frac{d}{dt}\int_\Omega E_m = \chi\int_\Omega q\nabla.(h\mathbf{u}) -
q^2(\nabla .\mathbf{u})/2
,
\end{align}

where $\Omega$ is any area over which the divergence is assumed zero.
$ E_m = (gm^2 + m\mathbf{u.u})/2$, where $m = h-\chi q$, a simplified
representation of the moist static energy.
\subsection{Asymptotic Analysis}

On a $\beta$-plane, we scale the equations as:
$
u,v \sim U(u,v),~ 
t \sim L/U,~  
h \sim H + P\eta$ and 
$\tilde{q},q_s \sim  Q_0$. 
Non-dimensionalising using the shallow water definitions of Rossby Number
($Ro = U/f_0L$), Froude Number ($Fr = U/\sqrt{gH}$), relative height of waves ($\Gamma =
P/H$), ratio of condensation time scale to advective time scale ($\alpha =
\tau_q/(L/U)$)
and strength of the $\beta$-effect ($r = \beta L/f_0$), (\ref{a1}) takes the form
\begin{align}
\frac{Du}{Dt} - \frac{1}{Ro}(v + rvy) = -\frac{\Gamma}{Fr^2}\eta_x, \nonumber \\
\frac{Dv}{Dt} + \frac{1}{Ro}(u + ruy)= -\frac{\Gamma}{Fr^2}\eta_y,  \nonumber \\
\frac{D\eta}{Dt} + \frac{1}{\Gamma}(\nabla .\mathbf{u}) + \eta(\nabla
.\mathbf{u}) = - \delta\frac{\chi Q_0 \tilde{q}}{P\alpha}, \nonumber \\
\frac{D\tilde{q}}{Dt} + \tilde{q}(\nabla .\mathbf{u}) + \nabla \cdot (\mathbf{u} q_s) = -\frac{\delta}{\alpha} \tilde{q}, 
\label{a2}
\end{align}
where $Q_0$ is the scale parameter for $\tilde{q}$. Taking the limit of small Rossby, Froude and perturbations
($Ro = Fr = \Gamma = r = \epsilon$), using a ``slow" mass source ($\alpha
= 1$) and assuming an $\mathcal O(1)$ contribution to the height field from
$\chi Q_0/P $, we get 
\begin{align}
\frac{Du}{Dt} - \frac{1}{\epsilon}(v + \epsilon vy) = -\frac{1}{\epsilon}\eta_x, \nonumber \\
\frac{Dv}{Dt} + \frac{1}{\epsilon}(u + \epsilon uy) = -\frac{1}{\epsilon}\eta_y, \nonumber  \\
\frac{D\eta}{Dt} + \frac{1}{\epsilon}(\nabla .\mathbf{u}) + \eta(\nabla
.\mathbf{u}) = - \delta \tilde{q}, \nonumber \\
\frac{D\tilde{q}}{Dt} + \tilde{q}(\nabla .\mathbf{u}) + \nabla \cdot (\mathbf{u} q_s) = -\delta \tilde{q}. 
\label{a3}
\end{align}
Expanding all fields in an asymptotic series in powers of $\epsilon$, and
collecting the dominant terms ($\mathcal O(1/\epsilon)$), we obtain the
following balance,
\begin{align}
v^0 = \eta^0_x, ~ 
u^0 = -\eta^0_y, ~ 
\nabla .\mathbf{u^0} = 0.
\label{a4}
\end{align}
Which is nothing but a non-divergent geostrophic flow. 
Note that, with $\chi Q_0/N \sim \mathcal O(1)$,  if we assume ``fast" condensation and use $\alpha = \epsilon$, we
end up with an inconsistent set of equations at $\mathcal O(1/\epsilon)$.
Thus, a ``slow" mass source is essential to
maintain geostrophy in the leading order fields. Also, this time scale is
independent of the functional form of the forcing, which need not be closed on
the vapour equation as has been done in our formulation.

Proceeding with the asymptotic expansion, at $\mathcal O(1)$,
we have,
\begin{align}
\frac{D^0u^0}{Dt} - v^1 - v^0y= -\eta^1_x, \nonumber \\
\frac{D^0v^0}{Dt} + u^1 + u^0y= -\eta^1_y, \nonumber \\
\frac{D^0\eta^0}{Dt} + (\nabla .\mathbf{u^1})  =
-\delta \tilde{q}^0, \nonumber \\
\frac{D^0 \tilde{q}^0}{Dt} +  \nabla \cdot (\mathbf{u}^0 q_s) = - \delta \tilde{q}^0. 
\label{a5}
\end{align}
Taking the curl of the first two equations and substituting for $\nabla .
\mathbf{u^1}$ from the third, we get (dropping superscripts as all fields are at order zero) a
``QG-like" vorticity equation coupled with 
moisture advection by the order zero velocity fields. Specifically,
\begin{align}
\frac{D}{Dt}(\nabla^2 \eta - \eta) + \eta_x - \delta \tilde{q} = 0, \nonumber \\
\frac{D \tilde{q}}{Dt} + (\mathbf{u} \cdot \nabla) q_s + \delta \tilde{q} = 0. 
\label{a6}
\end{align}
Equations (\ref{a6}) represents the ``slow" evolution of the moist rotating shallow water system (\ref{a1}). 
We emphasise that, in (\ref{a6}), the time scale at which the effects
of condensation manifest themselves is of the same order as 
the advective time scale. 

Understandably, most applications of a moist framework have been in
tropical settings, and in contrast to the present scenario, the time 
scale of condensation is usually taken to be much smaller than that of
advection. In fact, a limit that has proved useful in tropical
analysis is $\tau_q \rightarrow 0$, the so-called limit of rapid
condensation \citep[see for example,]
[for dynamic and kinematic applications, respectively]{friersonFronts2004, sukhatmeYoung2011}
or strict quasi-equilibrium (SQE) \citep{emanuel1994}.
Midlatitude studies which look at the effect of latent heat release on baroclinic waves generally use different moisture
    parameterisations: in particular, wave-CISK and large-scale rain schemes
    \citep{devriesBaroclinic2010}. Wave-CISK closes the heating
    on the QG vertical velocity ($w$) at some level (for example, the cloud base $z_b$) [i.e., $Q \propto w(z_B)h(z)$] where $h(z)$
    is the vertical profile of heating. Large scale rain schemes close
    the heating via $Q \propto w(z)r(z)$, where $r(z)$ is the
    vertical specific humidity profile. In both these cases, the vertical profile of
heating/humidity is important. Further, these schemes usually do not consider the effects of horizontal gradients of
moisture.

In the tropical context, the idea of using forcing due to condensation as a slow, second order
term in an asymptotic expansion has been explored by
\cite{yano2009}.
Furthermore, in the lifecycle of mesoscale convective systems,
\cite{mapes2006} observe significant moisture anomalies in the
free troposphere for as much as four days before and after peak rainfall, 
and suggest that the long duration of these anomalies might project mesoscale
variability onto
intraseasonal scales.
In a different context,
\cite{lindzen2003} showed that moist convective response time must exceed the
lifetime of a single cumulus cloud to account for the observed phase of the
semidiurnal tide. 
More broadly, ``slow" forcing of
the full SWE, that projects mainly on the balanced dynamics, has been
adopted in models that study the effect of moist convection on large
scale circulation in the atmospheres of Jupiter and Saturn
[\citep{showman2007}, see also the discussion in \cite{scottPolvaniSphericalSWE2007}].
In addition,
for slow, large scale modes such as the Madden-Julian Oscillation (MJO), studies have shown the
importance of the horizontal advection of moist static energy \citep[see, for
example,][]{maloneyMoist2009, andersenMoist2011, pritchardCausal2014, sobelMoist2014}. Also, convective parameterisations that are based
on relative humidity are seen to perform better than those based on convergence in simulating the
MJO \citep{hironsUnderstanding2013}.
In fact, a ``mechanism-denial'' experiment showed that the MJO shuts off in a GCM when the rotational
component of horizontal moist static energy advection is removed \citep{pritchardCausal2014}.
Thus, given the slow time scales of the rotational modes in the atmosphere, we argue that the main
features of our moist model, namely \textbf{a)} slow time scales, \textbf{b)}
non-divergent advection and \textbf{c)}
relative (column) humidity based latent heat release can be justified on physical grounds.

\subsection{Vapour Equation}

To facilitate study on a doubly periodic domain, we follow previous work 
and assume a linear gradient in the
meridional direction of the saturation vapour specific humidity
\citep{sobel2001}. In addition, based on
Figure \ref{pwatGradients}, we also consider a linear dependence of $q_s$ with longitude.
Even though the temperature may be fairly uniform in longitude, which
leads to a uniform $q_s$ from the Clausius-Clapeyron relation, it
should be remembered that this relation is an upper bound. Indeed, the
general circulation of the atmosphere gives character to the $q_s$
profile. 

Therefore, a simple prescription for the saturation profile reads, $q_s= Q_0 (1- \beta_q \frac{y}{L} - \alpha_q \frac{x}{L})$. 
As $\alpha_q$ can be positive or negative, we have $\alpha_q,\beta_q < 1$ individually,
and $\alpha_q+\beta_q < 1$ in combination to enforce positive-definiteness of
$q_s$. Due to the scaling of $x,y$ by $L$, $\alpha_q$
and $\beta_q$ are independent of the size of the domain; they are given by
$\Delta Q/Q_0$, where $\Delta Q$ is the change in water vapour along the zonal
and the meridional direction for $\alpha_q$ and $\beta_q$ respectively. A
domain with a high value of $Q_0$ (close to the tropics, for example) will have
smaller values of $\alpha_q,\beta_q$.
Scaling $q_s$ by the same factor as $\tilde{q}$, non-dimensionalising and substituting into (\ref{a6}), we get 
\begin{align}
\frac{D}{Dt}(\nabla^2 \eta - \eta) + \eta_x - \delta \tilde{q} = 0, \nonumber \\
\frac{D\tilde{q}}{Dt} = \beta_q v +\alpha_q u - \delta \tilde{q}.
\label{a7}
\end{align}
These equations form the basis of our analysis in the following sections.

\subsection{Linear Analysis}

Linearising (\ref{a7}) about a saturated state of rest, we obtain
\begin{align}
\frac{\partial}{\partial t}(\nabla^2 - 1)\eta + \eta_x = \delta \tilde{q},
    \nonumber \\
\frac{\partial \tilde{q}}{\partial t} - \beta_q\eta_x - \alpha_q(-\eta_y) = -\delta \tilde{q},
\label{b1}
\end{align}
where the geostrophic velocity has been expressed in terms of $\eta$. In the non-condensing phase, i.e., with $\delta=0$, the equations in (\ref{b1}) 
are uncoupled and
reduce to the passive advection of moisture by Rossby waves. With $\delta=1$,
the condensing phase, on Fourier transforming in space and time and re-arranging terms yields
\begin{align}
\left [ \begin{array}{cc} \frac{-k_x}{1+k_x^2+k_y^2} & \frac{-i}{1+k_x^2+k_y^2} \\
-k_x\beta_q+k_y \alpha_q & -i \end{array} \right ]
\left [ \begin{array}{c} \hat\eta \\ \hat{\tilde q}  \end{array}\right ] 
=\sigma\left [ \begin{array}{c} \hat\eta \\ \hat{\tilde{q}}  \end{array}\right ].
\label{b2}
\end{align}

Where $k_x,k_y$ are the spatial wavenumbers and $\sigma$ is the phase speed of
the wave. Fourier transformed quantities are represented with a hat.
In the absence of gradients, the matrix on the left hand side is upper
triangular, and the eigenvalues are given by the diagonal elements. Thus, the
eigenvalues are $\{-k_x/(1+k_x^2+k_y^2),-i\}$, which correspond to a dry
Rossby wave and an exponentially damped mode, respectively. The
eigenvector corresponding to the Rossby wave is of the form $e_R =
[c\hspace{5pt} 0]^T$ for
arbitrary real-valued $c$, indicating a decoupling of the Rossby wave from the
effects of condensation. Thus, 
this analysis immediately points to the crucial role of moisture gradients in the present non-divergent framework.

\subsubsection{Purely meridional gradients ($\alpha_q=0, \beta_q \neq 0$)}

We begin by considering only meridional gradients in the saturation profile, i.e., $\beta_q \neq 0, \alpha_q = 0$.
Taking the limit of large $k_x$ in (\ref{b2}), we see that in both rows
of the matrix on the LHS, the term in
the second column is small as compared the term in the first column,
and hence the effects of condensation are minimal. Therefore, for strong coupling between 
moisture and $\eta$, much of our attention is restricted 
modes with small values of $k_x$, i.e,
those with a large zonal scale.

\begin{figure}
\centering
\includegraphics[width=0.45\textwidth]{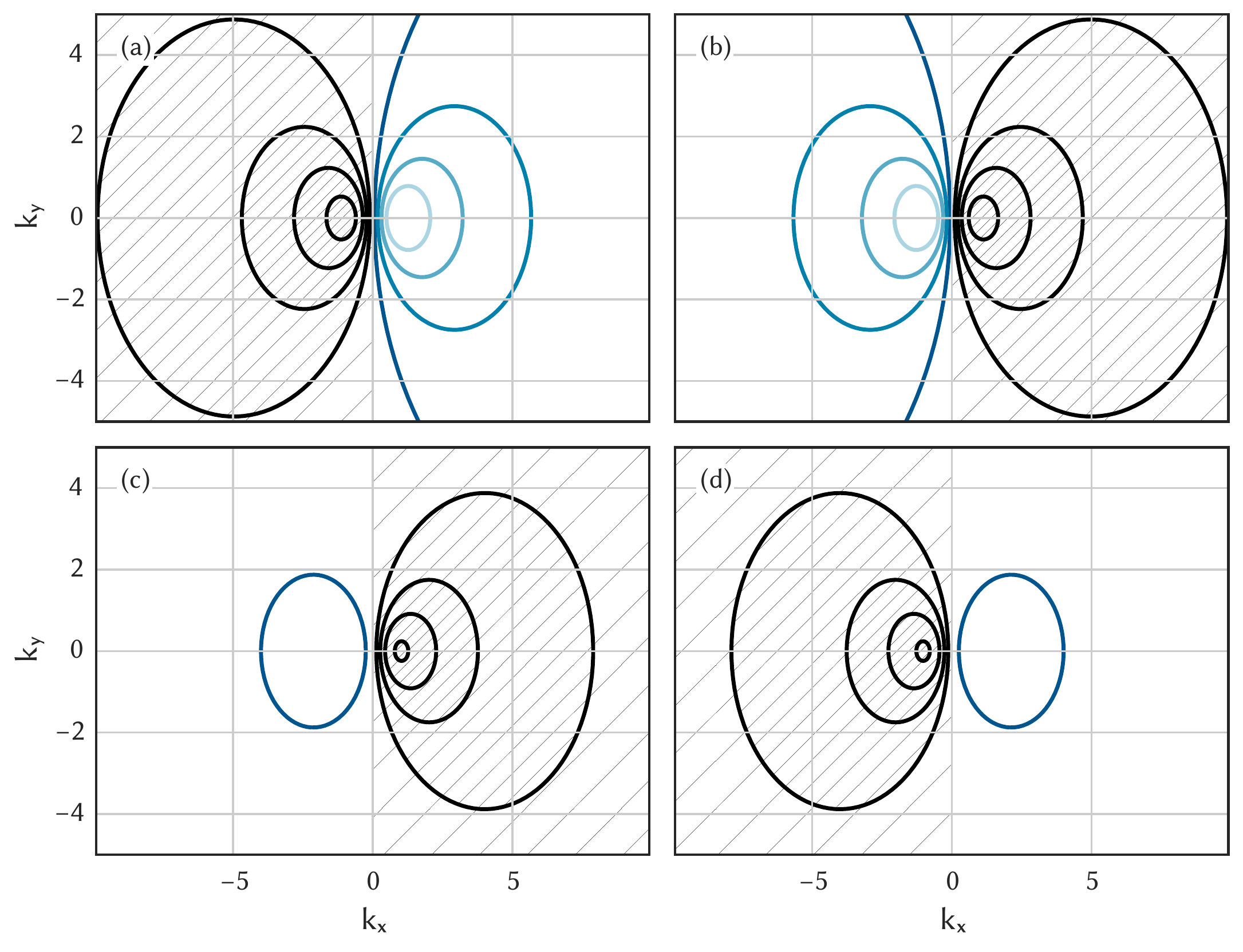}
\caption{Phase speed ($\Re(\sigma)$) corresponding to $\alpha_q=0$ and $\beta_q=0.2$ is
    plotted in panels \textbf{a,b} and $\beta_q=0.8$ in panels
\textbf{c,d}. The left panels correspond to eigenvector 1 and the right panels correspond to
eigenvector 2. The region of the $k_x-k_y$ plane where the modes are strongly damped ($\Im(\sigma) <
-0.1$) is hatched.} 
\label{dispRelonlyBeta}
\end{figure}

\begin{figure}
\centering
\includegraphics[width=0.45\textwidth]{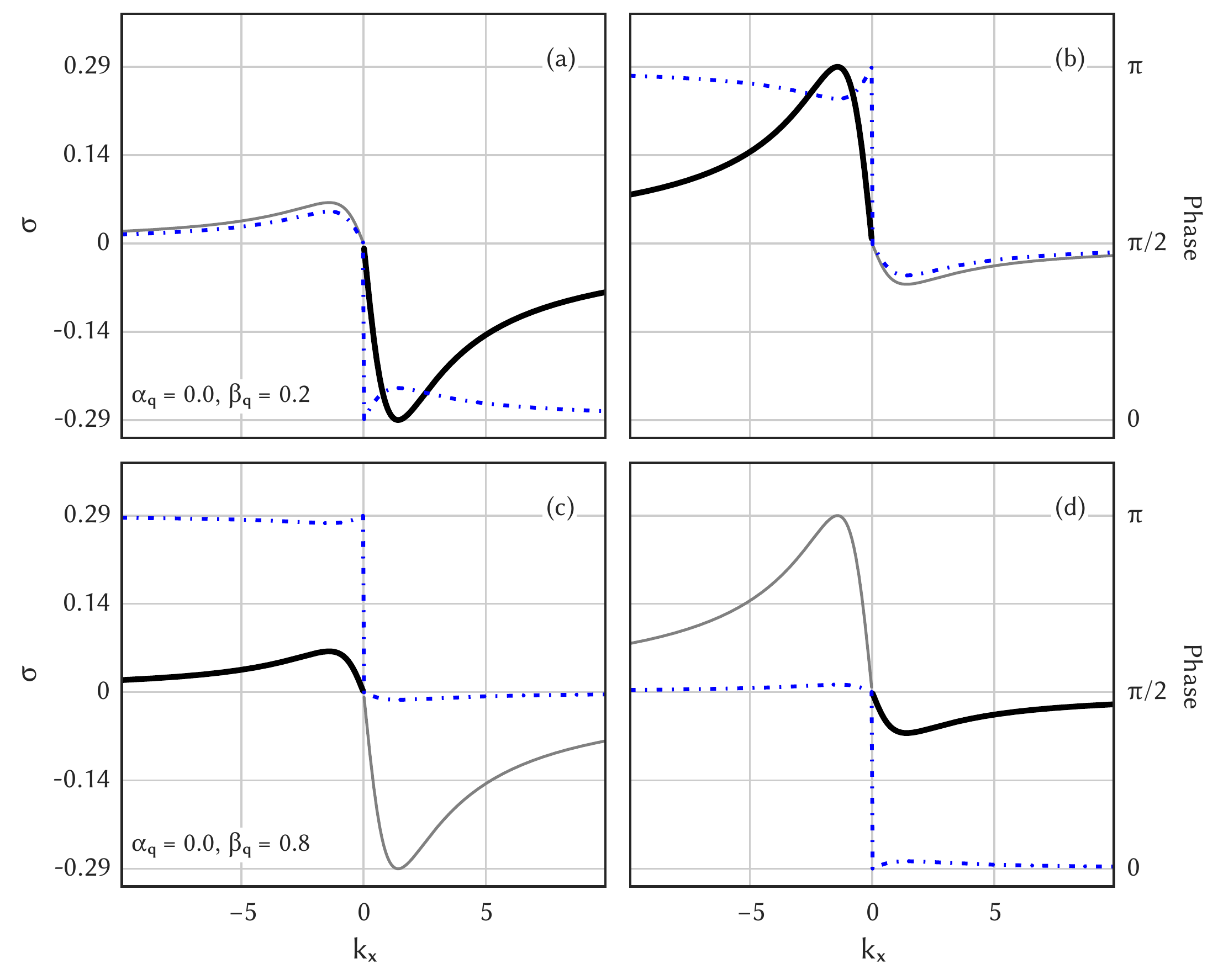}
\caption{ Phase speed ($\Re(\sigma)$) and phase relationship between $v$ and $\tilde
    q$ for $k_y=2$. Panels \textbf{a,b} correspond to $\alpha_q,\beta_q = 0.0,0.2$ and
        \textbf{c,d} correspond to $\alpha_q,\beta_q = 0.0,0.8$.
        $\Re(\sigma)$ is plotted in thick black and thin gray lines for the almost neutral
and strongly damped modes respectively. 
The phase difference is plotted in
blue dash-dot lines.} 
\label{1dOnlyBeta}
\end{figure}

\begin{figure}
\centering
\includegraphics[width=0.45\textwidth]{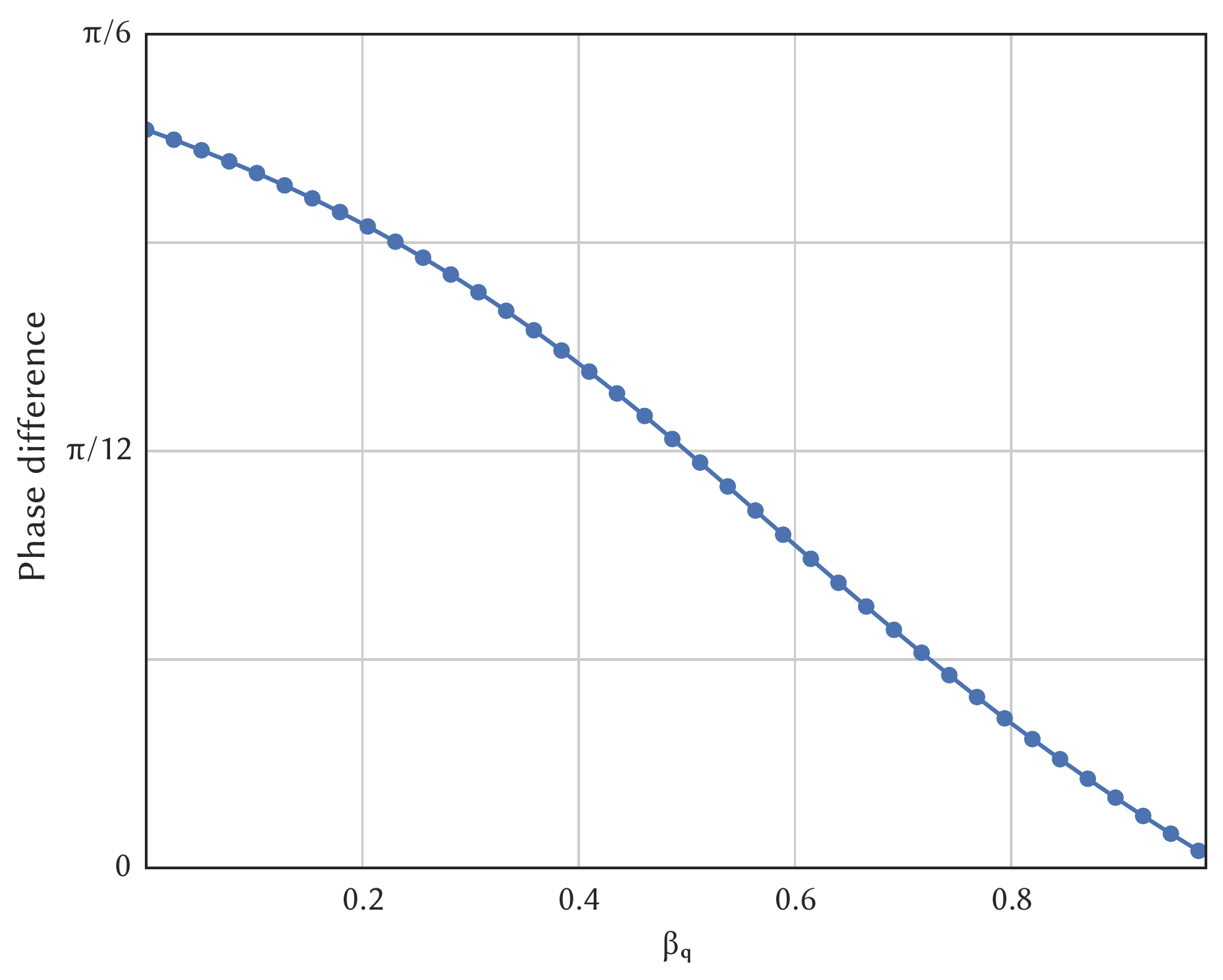}
\caption{The variation of the maximum of the phase difference between $v$ and $\tilde{q}$ with $\beta_q$.} 
\label{phaseVariation}
\end{figure}

The dispersion relation for two representative cases ($\beta_q=0.2,0.8$) is shown in Figure \ref{dispRelonlyBeta}.
The differences between these moist waves and dry Rossby waves are noted below:
\begin{itemize}
\item
One mode is very weakly damped (almost neutral),
while the other is
strongly damped (with $\mathcal{O}(1)$ damping). In particular, waves associated with the
first eigenvector are strongly damped for $k_x < 0$ and those associated with the
second eigenvector are strongly damped for $k_x > 0$ for $\beta_q < 0.5$. This situation is reversed
when $\beta_q>0.5$. 
\item
The corresponding phase
relationship between the meridional velocity $v$ and $\tilde q$ is shown in Figure \ref{1dOnlyBeta} 
for $\beta_q = 0.2,0.8$ respectively.
For the strongly damped waves, the meridional velocity and condensation are in quadrature. Thus, the
condensation occurs closer to the centre of the Rossby gyre rather than the flanks, this appears to
strongly damp the wave. For
the almost neutral wave, the meridional velocity and condensation are nearly in phase. Thus, the condensation is
co-incident with the meridional velocities at the flanks of the Rossby gyres which results in a marginal damping, 
but slows down the wave.
\item
 Figure \ref{1dOnlyBeta} also shows the dispersion relation (for a fixed $k_y$). Clearly, as $\beta_q$ increases, the frequency 
 corresponding to the neutral modes decreases, i.e., the phase speed decreases with a stronger gradient \citep[consistent with][]{sukhatmeMoistModes2013}. 
Further, as reported, $v$ and $\tilde q$ are almost perfectly in phase. 
Note that we are referring to the phase difference for small $k_x$, and not the 
asymptotic value as $k_x$ becomes large. In fact, as $\beta_q \rightarrow 1$, the 
absolute mimimum phase difference (around $|k_x| \approx 2$) tends to zero (Figure
\ref{phaseVariation}). This is also the case when $k_y$ is
increased keeping $\beta_q$ constant (not shown).
\item
As $\beta_q$ decreases, the absolute mimimum phase difference between $v$ and $\tilde
q$ increases from zero, and reaches a minimum value between $\pi/4$ and
$\pi/2$ for the fastest modes as shown in Figure \ref{phaseVariation}.
Concomitantly, the frequency of the moist waves approaches that of dry
Rossby waves. 
\end{itemize}

The decrease in phase speed with increasing $\beta_q$ can be qualitatively understood by examining
(\ref{b1}). Specifically, since $v$ and $\tilde q$ have the same
    phase as $\beta_q \rightarrow 1$ the tendency term in the PV equation becomes small and results in a slowly
or almost non-propagating wave.
More clearly, the reduction in phase speed of the moist modes in the presence of condensation
can also be understood
by linearising the potential vorticity (PV) of the
full moist shallow water equations. For a state of rest, the moist PV is
given by:
\begin{align}
\textrm{PV} &= \frac{f_0 + \beta y}{H - \chi Q_0(1-\beta_q y)}, \\\nonumber
    &=\frac{f_0 + \beta y}{H_q (1 + M_\beta y)}, \\\nonumber
    &\approx (f_0 + \beta y)(1 - M_\beta y)/H_q, \\\nonumber
    &\approx (f_0 + (\beta - f_0M_\beta)y)/H_q, \\\nonumber
\end{align}
where $H_q = H_0 - \chi Q_0$ and $M_\beta = \beta_q\chi Q_0/H_q$.
Thus, we see that the moisture gradient counters the effect of the dynamical
$\beta$, similar to the effect of a sloping bottom topography. This implies
that the advection of planetary vorticity by the winds generated by the
propagation of the Rossby wave is counteracted by condensation/drying occuring due to
the same winds. Therefore, there is a smaller net increase/decrease in PV,
resulting in slower Rossby wave propagation. We note that if $\beta_q>1$, the
planetary vorticity advection is overwhelmed by the condensation, and an
eastward moving unstable mode results. However, in our system we require $\beta_q\le 1$, and
therefore this situation does not occur.

\subsubsection{Purely zonal gradients ($\alpha_q \neq 0, \beta_q = 0$)}

\begin{figure}
\centering
\includegraphics[width=0.45\textwidth]{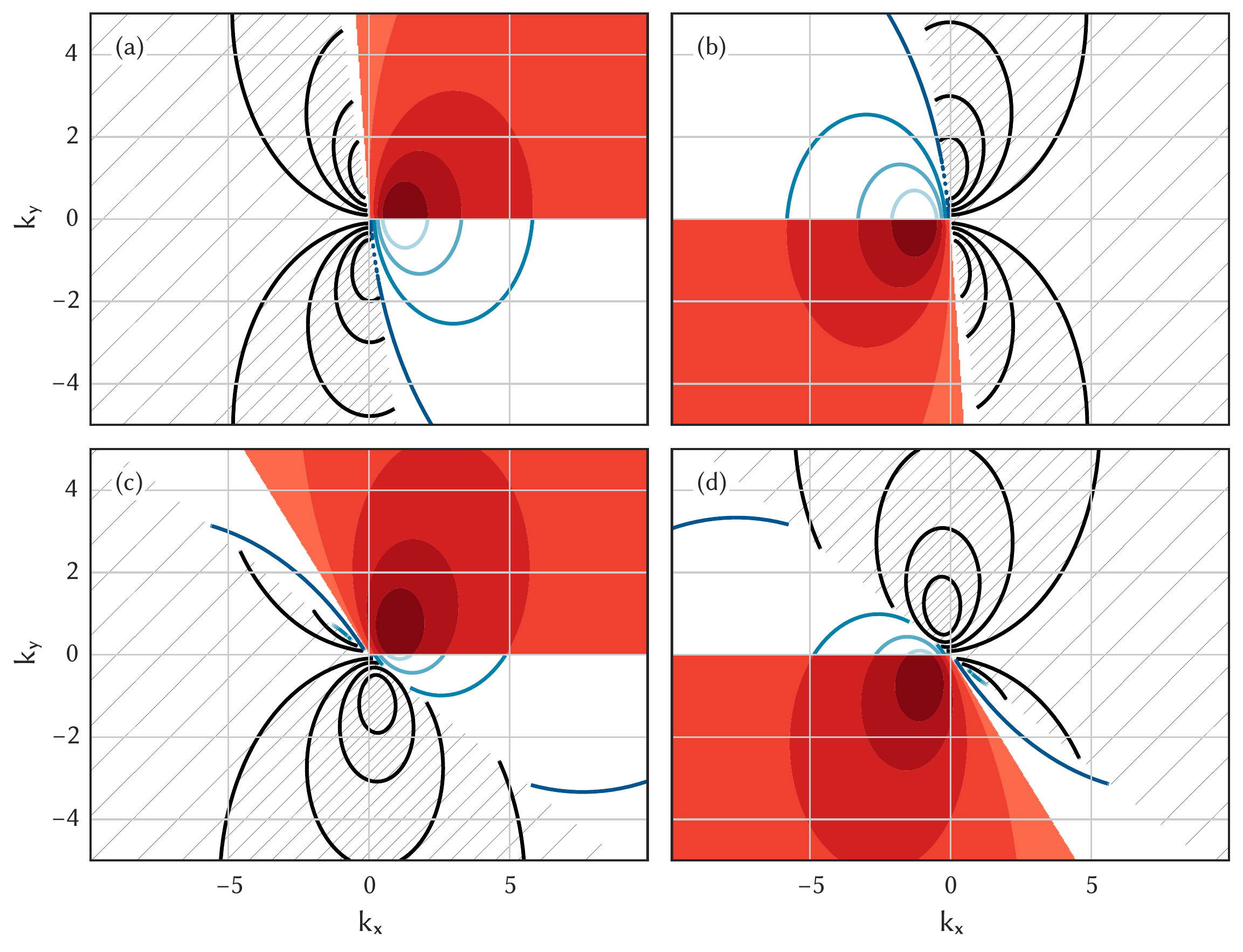}
\caption{ Phase speed ($\Re(\sigma)$) corresponding to $\beta_q=0$ and
$\alpha_q=0.1$ is plotted in panels \textbf{a,b} and $\alpha_q=0.9$ in panels
\textbf{c,d}.  The left panels correspond to eigenvector 1 and the right panels correspond to
eigenvector 2.  The region of the $k_x-k_y$ plane where the modes are strongly damped ($\Im(\sigma) <
-0.1$) is hatched.  The region of the $k_x-k_y$ plane where the modes are unstable ($\Im(\sigma) >
0$) is shaded red.
} 
\label{dispRelOnlyAlpha}
\end{figure}

\begin{figure}
\centering
\includegraphics[width=0.45\textwidth]{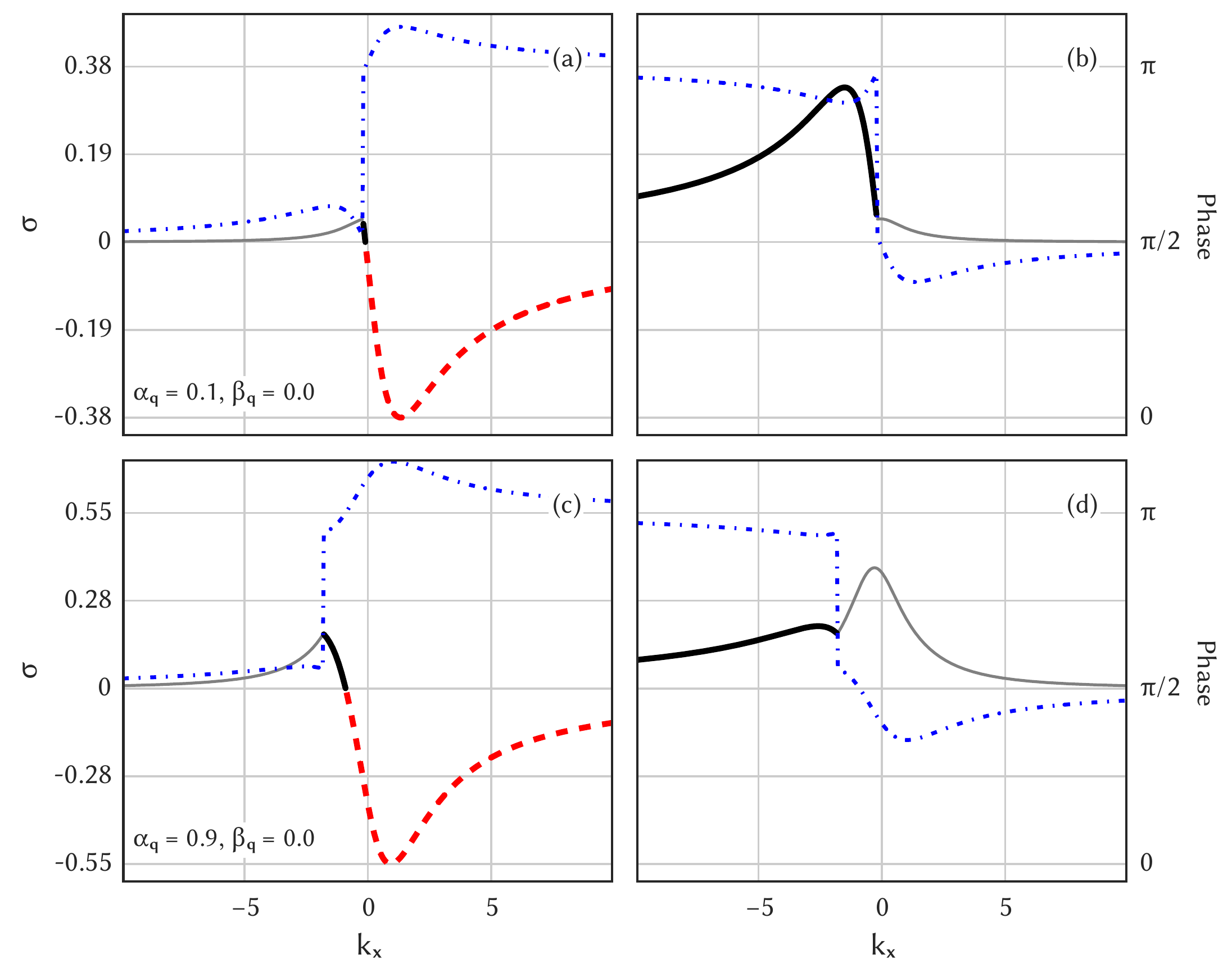}
\caption{Phase speed ($\Re(\sigma)$) and phase relationship between $v$ and $\tilde
    q$ for $k_y=2$. Panels \textbf{a,b} correspond to $\alpha_q,\beta_q = 0.1,0.0$ and
        \textbf{c,d} correspond to $\alpha_q,\beta_q = 0.9,0.0$.
        Mildly damped, strongly damped and unstable modes are plotted in bold black,
        thin grey and dashed red lines respectively.
The phase difference is plotted in
blue dash-dot lines.} 
\label{1dOnlyAlpha}
\end{figure}

The dispersion relation for $\alpha_q = 0.1,0.9$ are shown in Figure
\ref{dispRelOnlyAlpha}. Note that our notation is such that a
positive $\alpha_q$ corresponds to a decrease in the saturation vapour to the
east and vice versa. The differences from the previous case with only a meridional gradient are noted below:
\begin{itemize}
\item
The inclusion of a zonal gradient results in an instability (marked by the filled contours in Figure \ref{dispRelOnlyAlpha}), and the area in the
wavenumber space covered by the instability increases with increasing
$\alpha_q$. Further, inspecting Figure \ref{dispRelOnlyAlpha}, we note that for small 
$\alpha_q$ the unstable modes propagate westward, while for large $\alpha_q$ there emerges an
eastward propagating unstable mode with a large zonal scale.
\item
The phase relation in Figure \ref{1dOnlyAlpha}\textbf{a,c} suggests that both westward and eastward unstable modes are 
characterised by $v$ leading $\tilde q$ by a value
greater than $\pi$. This is in contrast to a lag in the
previous case (with $\alpha_q=0, \beta_q \neq 0$).
\item
 Examining Figure \ref{1dOnlyAlpha}, we note that 
the dispersion relation of the unstable mode resembles that of a dry
Rossby wave, but is shifted along the $k_x$-axis. This shift is more pronounced as $\alpha_q$ increases (seen by 
comparing panels \textbf{a} and \textbf{c} of Figure \ref{1dOnlyAlpha}).
\item
When the sign of $\alpha_q$ is reversed, the
dispersion relation is reflected about the line $k_x = 0$ (not shown).
Therefore, the direction of propagation in the zonal direction remains unchanged
and is reversed in the meridional direction.
\end{itemize}

At a phase lag greater than $\pi$, the tendency terms are such that a positive (negative) value of
PV is associated with a positive (negative) tendency, and therefore the PV
field grows with time. The reversal in the meridional direction of propagation on changing the sign of $\alpha_q$
 can be explained by noting that the zonal geostrophic winds
are maximum on the northward and southward flanks of a Rossby ridge/trough.

\subsubsection{Zonal and meridional gradients ($\alpha_q, \beta_q \neq 0$)}

\begin{figure}
\centering
\includegraphics[width=0.45\textwidth]{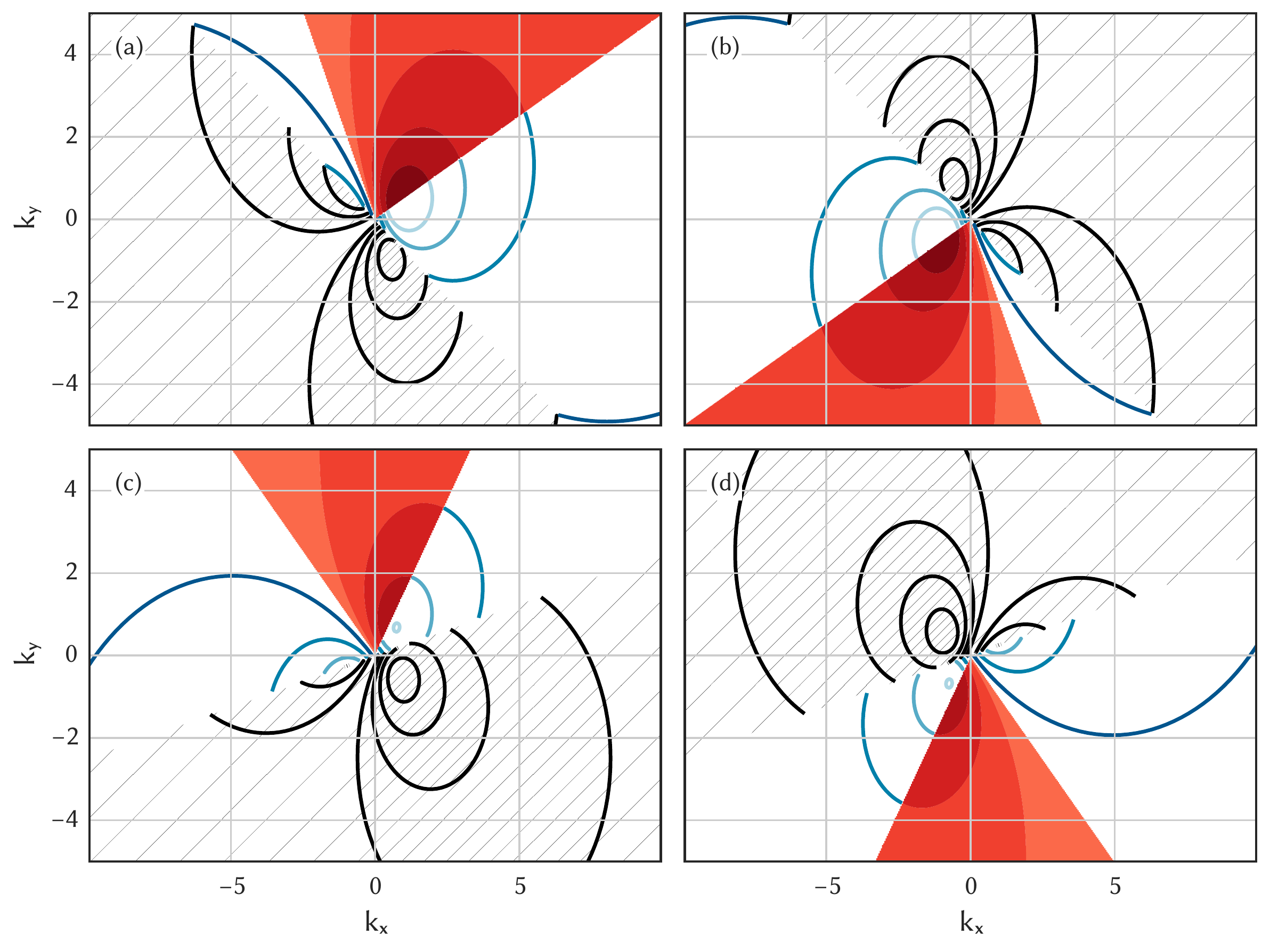}
\caption{ Phase speed($\Re(\sigma)$) corresponding to 
$\alpha_q,\beta_q=0.4,0.2$ is plotted in panels \textbf{a,b} and $\alpha_q,\beta_q=0.4,0.6$ in panels
\textbf{c,d}. 
The left panels correspond to eigenvector 1 and the right panels correspond to
eigenvector 2.  The region of the $k_x-k_y$ plane where the modes are strongly damped ($\Im(\sigma) <
-0.1$) is hatched.  The region of the $k_x-k_y$ plane where the modes are unstable ($\Im(\sigma) >
0$) is shaded red.
} 
\label{dispRelBothAlphaBeta}
\end{figure}

In a situation where both gradients are present, the behaviour is a mixture of
the two previously considered cases. As seen in Figure \ref{dispRelBothAlphaBeta}, if $\beta_q$ is increased while keeping $\alpha_q$
constant, the range of wavenumbers for which the unstable waves are eastward increases, while the
overall range of wavenumbers that are unstable is reduced. Further, there is a change in the direction of propagation of the most unstable modes with 
both $\alpha_q$ and $\beta_q$. Specifically, holding $\alpha_q$ fixed, the most unstable mode lies in the second and fourth
quadrants of Figure \ref{dispRelBothAlphaBeta}\textbf{a,b} and in the first and third quadrants of 
Figure \ref{dispRelBothAlphaBeta}\textbf{c,d}. Thus, 
for $\beta_q=0.2$ ($\beta_q=0.6$) 
the most unstable mode is eastward (westward), with the change occuring at $\beta_q=0.5$. Similarly, holding $\beta_q$ fixed, 
the direction of 
propagation reverses with a change in sign of $\alpha_q$ (not shown).
Quantitatively, the maximum non-dimensional growth rate is $\approx 0.04$
for $(\alpha_q,\beta_q=0.4,0.2)$ and $\approx 0.035$ for $(\alpha_q,\beta_q=0.4,0.6)$, which,
with an advective timescale of the order of a day or two, implies these modes mature in 25-50 days. Further, in both cases, these are very large scale 
modes whose dimensional phase speed is $\approx 1-3$ m/s.

As seen in Figure
\ref{pwatGradients}, and from observations in the tropics \citep{sukhatme2012},
the atmosphere is characterised by small values of $\alpha_q$ and relatively
larger values of $\beta_q$, with $\beta_q$ ranging between 0.5 and 0.7 and $\alpha_q$ ranging
between 0.3 and 0.4. From the above discussion on the direction of propagation
of the most unstable mode, noting that $\beta_q > 0.5$, this
implies that in regions with a negative (positive) $\alpha_q$, i.e, moisture 
 increasing (decreasing) towards the east, the most unstable modes are
eastward (westward). Thus, as was noted in a tropical setting \citep{sukhatmeMoistModes2013}, the modes grow in the direction of increasing
moisture. 

\begin{figure}
\centering
\includegraphics[width=0.45\textwidth]{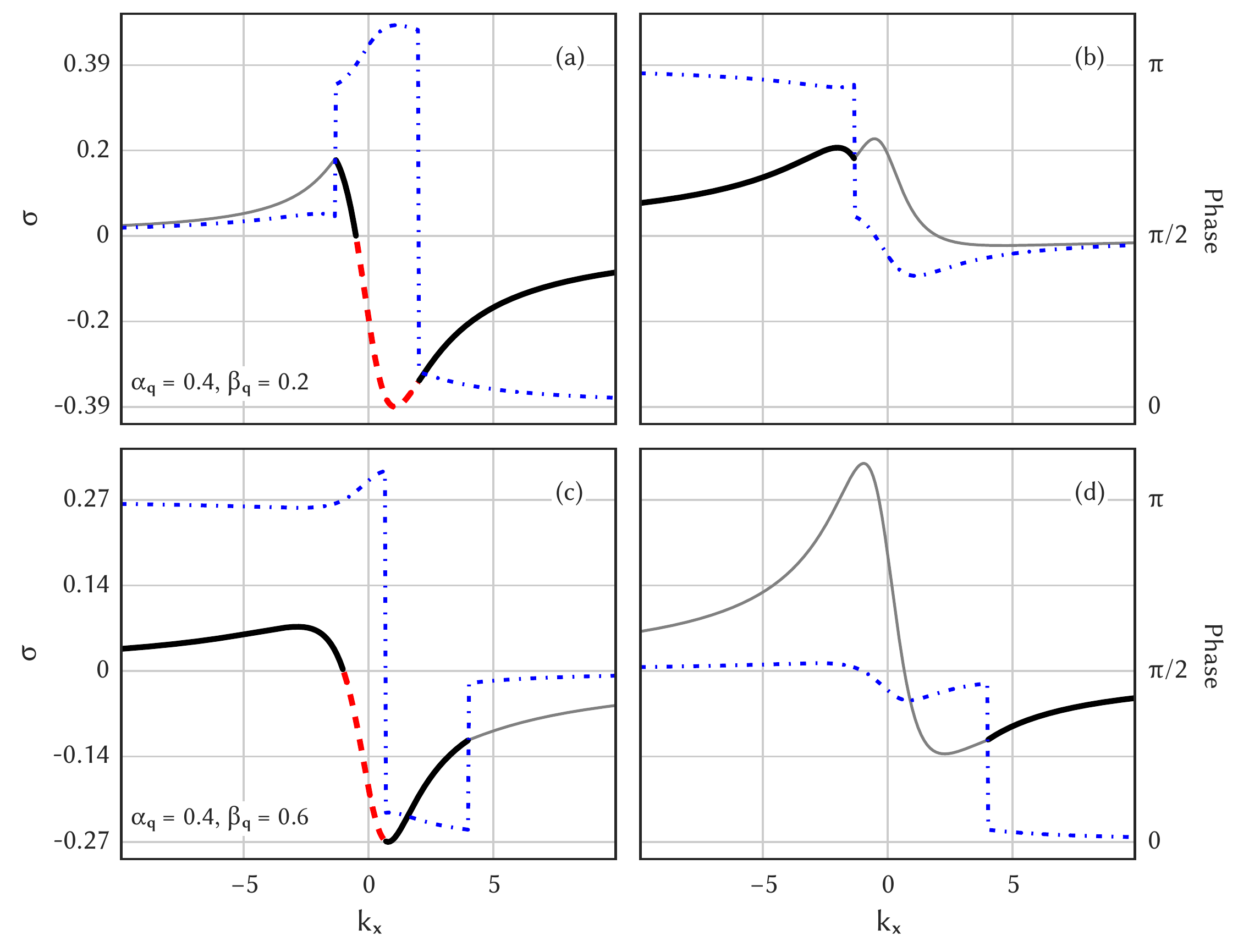}
\caption{Phase speed ($\Re(\sigma)$) and phase relationship between $v$ and $\tilde
    q$ for $k_y=2$. Panels \textbf{a,b} correspond to $\alpha_q,\beta_q = 0.4,0.2$ and
        \textbf{c,d} correspond to $\alpha_q,\beta_q = 0.4,0.6$.
        Mildly damped, strongly damped and unstable modes are plotted in bold black,
        thin grey and dashed red lines respectively.
The phase difference is plotted in
blue dash-dot lines.} 
\label{1dBothAlphaBeta}
\end{figure}


\subsubsection{The mechanism of the instability}

In our nondivergent system, there are three terms that play a role in the balance of PV.
Namely, the advection of planetary vorticity, condensation in the meridional
direction and condensation in the zonal direction. 
This is made clear by looking at the linearised PV with both
gradients present:
\begin{align}
\textrm{PV} &= \frac{f_0 + \beta y}{H - \chi Q_0(1-\beta_q y -\alpha_q x)}, \\\nonumber
    &=\frac{f_0 + \beta y}{H_q (1 + M_\beta y + M_\alpha x)}, \\\nonumber
    &\approx (f_0 + \beta y)(1 - M_\beta y - M_\alpha x)/H_q, \\\nonumber
    &\approx (f_0 + (\beta - f_0M_\beta)y - f_0M_\alpha x)/H_q, 
\end{align}
where $H_q=H_0- \chi Q_0$, $M_\alpha=\alpha_q \chi Q_0/H_q$ and $M_\beta=\beta_q\chi Q_0/H_q$.

In the meridional direction,
condensation and vorticity advection have opposing tendencies, and thus a larger
$\beta_q$ leads to a reduction in phase speed. By definition, $\beta_q \le 1$
and so we do not experience an instability. In the zonal direction, however, there is no
compensating term to balance condensation. Hence, the introduction of a nonzero gradient in the zonal direction 
immediately results in the generation of 
unstable modes. 
Similarly, with gradients in both directions, the instability arises at certain combinations of the moisture
gradients where the contribution to the PV field by the
dynamical $\beta$ is smaller than the contribution by the condensation.



\section{Summary and Discussion}

In this article, we have presented the derivation of a forced
QG system that highlights constraints placed on
the forcing function that are required to maintain geostrophic balance at leading order. 
The forcing is a due to a
dynamically active scalar which mimics the action of condensation,
thus providing a simple model for the moist atmosphere. We emphasize that the
constraints derived here --- in particular, the slow nature of the timescale of condensation --- have been used in previous 
studies of forced QG
and shallow water turbulence \citep{showman2007}, but normally without a formal 
justification.

Given that the time scale of condensation turns out to be 
comparable to that of advection, moisture anomalies are not
immediately relaxed to zero, as is the case with models that enforce SQE. 
Further, in the present model, heating is completely driven by non-divergent advection. 
In the presence of purely meridional moisture gradients, we only obtain damped and neutral modes. The neutral moist waves are 
noted to be slower than 
dry Rossby waves due to a reduction in an ``equivalent gradient" of PV. This is in contrast to SQE-based systems where 
a reduction in the equivalent depth is the cause for slower
convectively coupled waves \citep{kiladisCCEW}.
Further, the addition of meridional gradients on a $\beta$-plane complements the
work of \cite{sobel2001} by showing that the unstable
modes on a $f$-plane are stabilised in the presence of a planetary vorticity gradient.

This idea of ``diabatic PV gradients'' has been previously explored in
\cite[]{snyderQuasigeostrophic1991, parkerConditional1995} while studying the effects of latent
heat release on baroclinic wave development. In both these cases, Rossby waves are generated purely
due to diabatic processes and propagate via the advection of moisture and the
resultant heating by wind anomalies. This, is in some sense analogous to the results
of \cite{sobel2001}. The inclusion of boundaries moderates the instability, as suggested in
\cite{snyderQuasigeostrophic1991} and proved mathematically in \cite{devriesBaroclinic2010}. This is
due to the influence of the Rossby edge wave, which provides an opposing/stabilising
influence.
 These studies provide an intriguing parallel to our own in terms of the presence of
stabilising/destabilising PV gradients, though it is unclear whether they can be reduced to a
similar mathematical structure.

The inclusion of gradients in the zonal direction, by themselves or in combination with meridional 
gradients, is seen to result in the formation of unstable modes. The nature of the instability is deduced from the PV 
equation; specifically, an unstable situation occurs when the contribution to PV
from condensation overwhelms the stabilising 
dynamical $\beta$ contribution. It is seen that both zonal and meridional gradients affect the direction of propagation of the most unstable 
mode, In particular, for gradients that 
roughly correspond to atmospheric observations, the most unstable mode propagates zonally in the direction of increasing 
moisture. Further, the most unstable mode has a large spatial scale, small phase speed (of the order of a few m/s) 
and matures over intraseasonal timescales.

The limitations of the study pertain mainly to the geometry of the
domain under consideration and the use of a linear gradient in
$Q_s$ which enables us to obtain analytical results. To facilitate analysis in
the Fourier domain, we use $\delta=1$ throughout, which is not representative of
geophysical situations where precipitating and non-precipitating regions often
co-exist. However, we note that the mechanism in terms of the PV budget of the waves
    seems robust enough to apply in more general situations. Indeed, slowdown of Rossby waves is seen in the
eigenanalysis on an equatorial $\beta$ with meridional moisture gradients
\citep{sukhatmeMoistModes2013}, and preliminary numerical work on the QG system shows a slowdown of
the Rossby waves even in the presence of non-precipitating regions.
A study using spherical
geometry with more realistic gradients would also be interesting from a physical
perspective. Similarly, the analysis of the two layer case with baroclinic instability
and the interaction of Rossby waves which advect both PV and moisture appear to be fruitful avenues
of study. 

{\it Acknowledgements:} The authors thank the two reviewers and the editor for comments
which greatly added to the content and presentation of this work, especially in relating our work to
extant work on diabatic Rossby waves.
We thank Raghu Murtugudde for helpful
discussions. We thank Dargan Frierson whose suggestions helped guide the
approach and presentation of the paper. We thank Mike Wallace for
discussions that helped us gain physical insight into the problem, enabling
us to sketch out the underlying mechanisms.

\bibliography{moistQG}

\end{document}